\documentclass{aa}

\usepackage{graphics}

\newcommand{\bp}{$\beta\:$Pictoris}
\newcommand{\kms}{km\,s$^{-1}$}
\newcommand{\caii}{\mbox{Ca~{\sc ii}}}

\newcommand{\feii}{\mbox{Fe~{\sc ii}}}
\newcommand{\crii}{\mbox{Cr~{\sc ii}}}
\newcommand{\siii}{\mbox{Si~{\sc ii}}}
\newcommand{\niii}{\mbox{Ni~{\sc ii}}}
\newcommand{\mgii}{\mbox{Mg~{\sc ii}}}
\newcommand{\ci}{\mbox{C~{\sc i}}}
\newcommand{\si}{\mbox{S~{\sc i}}}
\newcommand{\erg}{erg~cm$^{-2}$~s$^{-1}$~\AA$^{-1}$}

\begin{document}
\thesaurus{08 (08.09.2 $\beta\:$Pic; 08.03.4; 08.16.2; 03.13.2)}

\headnote{Research Note}

\title{Possible spectral lines from the gaseous $\beta\:$Pictoris disk}

\author{
A. Lecavelier des Etangs \inst{1}
\and
L.M. Hobbs \inst{2}
\and
A. Vidal-Madjar \inst{1}
\and 
H. Beust \inst{3}
\and 
P.D. Feldman \inst{4}
\and 
R. Ferlet \inst{1}
\and 
A.-M. Lagrange \inst{3}
\and 
W. Moos \inst{4}
\and 
M. McGrath \inst{5}
}
        
\institute{
Institut d'Astrophysique de Paris, CNRS, 98 bis Bld Arago, F-75014 Paris, France
\and
Yerkes Observatory, University of Chicago, Williams Bay, WI 53191-0258, USA
\and
Laboratoire d'Astrophysique, Universit\'e J. Fourier, BP 53, 
F-38041 Grenoble Cedex 9, France
\and
Department of Physics and Astronomy, Johns Hopkins University, Baltimore,
Maryland 21218, USA
\and
Space Telescope Science Institute, 3700 San Martin Drive, Baltimore, 
MD 21218, USA
}

\date{submitted to Astronomy \& Astrophysics}
\date{Received / Accepted}

\maketitle

\begin{abstract}

We present the results of exploratory HST observations 
to
detect the emission lines
of ions in the \bp\ disk with the spectrograph slit placed off the star.
Possible emission lines from \feii\ have been detected at 
$\sim 10^{-14}$erg~cm$^{-2}$~s$^{-1}$ at 0.5\arcsec\ from \bp, 
which would suggest a total \feii\ density of 
$\ga 2\cdot 10^{-2}$cm$^{-3}$ at 10~AU. 
This detection 
needs confirmation.
If real, it
requires a
large production rate of gas and dust 
equivalent to the total disruption of 
ten bodies 30 kilometers in radius per year.


\end{abstract}

\section{Introduction}
\label{intro}

The \bp\ disk is an evolved replenished circumstellar disk around a 
main sequence star, and appears to be a planetary (or cometary)
disk (Vidal-Madjar et al. 1998).
In addition to the 
dust 
disk, 
gas has been probed through absorption lines 
on the star line of sight (Lagrange et al. 1998).
The observations of this gas component historically gave the most 
unexpected and important results with the detection of what
appears to be the first extra-solar comets ever observed
(Ferlet et al. 1987, Vidal-Madjar et al. 1998).
However, non-radial motions are still unknown because observations are 
limited to the gaseous absorption lines on the star line of sight.
Here we present the results of exploratory HST observations 
aimed to observe the emission lines
of ions in the disk with the spectrograph slit 
off the star.

\section{Observations and data analysis}

\subsection{Observations}
\label{obs}
\label{data}

The observations were made on February 9, 1996 with the HST/GHRS using
the SSA slit (0.22\arcsec$\times$0.22\arcsec ) and echelle spectroscopy
at an expected resolution of 80~000
to get the highest chance of detection of lines with {\it a priori}
unknown width. 
After a peak-up on the star \bp\
and the acquisition of reference spectra, the telescope was
shifted to the south-west part of the \bp\ disk at a position angle
of 31.5 degrees to a distance of 0.5 and 1.5\arcsec\ from the
star. The log of the observations is summarized in Table~\ref{log}.
The candidate lines are those from ions for which strong absorptions have been
seen in the stable component due to the gas disk and in the 
variable components due to the infalling comets.
On this basis, we expected the strongest fluorescence.

\begin{table}[bht]
\caption{Log of the observations performed on Feb. 9$^{\rm th}$ 1996.
$r$ is the distance from \bp }
\label{log}
\begin{tabular}{|c|c|clll|}
\hline
\#&$r$ & Wavelength & Lines & Time & $T_{exp}$  \\
&& (\AA) && (UT) & (s) \\
\hline
1&0.0 \arcsec & 1803.1 - 1812.7 & \niii & 08:39:03& 230\\
 &            &                 &  \si, \siii      &  & \\
2&0.0 \arcsec & 2373.0 - 2384.6 & \feii & 08:46:42 & 173\\
\hline
3&0.5 \arcsec & 2373.0 - 2384.6 & \feii & 08:55:45 & 2016\\
4&0.5 \arcsec & 1803.1 - 1812.7 & \niii & 10:28:45 & 2073\\
 &            &                 &  \si, \siii     &  & \\
\hline
5&1.5 \arcsec & 1803.1 - 1812.7 & \niii  & 11:58:39& 3571\\
 &            &                 &  \si, \siii     &  & \\
\hline
6&0.5 \arcsec & 1652.7 - 1661.2 & \ci  & 14:46:39 &2534 \\
 &            &                 & \ci$^*$, \ci$^{**}$     &  & \\

\hline
\end{tabular}
\end{table}

\subsection{The method}
\label{method}
\label{si spectrum at 0.5}

The spectra observed on the disk present the same general shape as
the spectra on the star but with a lower intensity level.
Indeed, any spectrum of the disk obtained off the star ($F_{obs}(\lambda)$)
is an addition of a ``noise'' component and a spectrum carrying information 
from the disk at the pointed position.
The ``noise'' is the spectrum of the starlight scattered by the telescope 
($a F_*(\lambda)$, where $F_*(\lambda)$ is the star spectrum)
plus a background level ($b$) 
due to the fact that the background determined by the HST pipeline can 
be slightly miscalculated. 
The part of the spectrum due to the disk ($F_{disk}$) is 
the starlight scattered 
by the dust and the possible emission lines due to gaseous fluorescence.
We thus consider that the additional noisy component
($F_{noise}$) is a linear combination of the stellar spectra ($F_*(\lambda)$) 
obtained at the same wavelength a few orbits before:
\begin{equation}
F_{obs}(\lambda)= F_{noise}(\lambda) + F_{disk}(\lambda) 
\approx a F_*(\lambda) + b + F_{disk}(\lambda)
\end{equation}
To put forward the potential presence of a component different from the 
dominant 
starlight scattered by the telescope, 
we divide the observed spectra by the stellar spectra.
The presence of a disk contribution at a wavelength $\lambda$ 
will be detected if the ratio 
$F_{obs}(\lambda)/(a F_*(\lambda) + b)$ is statistically significantly different from 1.

The main problem is thus the determination of $a$ and $b$, 
and their errorbars within a confidence 
level. Assuming that the disk spectra (dust scattered light plus gas emission 
lines) have a negligible contribution to the observed spectra
($F_{noise}(\lambda) \gg  F_{disk}(\lambda)$), we can determine
$a$ and $b$ by a $\chi^2$ minimization of 
\begin{equation}
\chi ^2 = \sum _{\lambda_i} w_{\lambda_i} 
  (a F_*(\lambda_i) + b - F_{obs}(\lambda_i))^2 ,
\end{equation}
where $w_{\lambda_i}= 1/\sigma_{\lambda_i}^2$
is the weight of each measurement at $\lambda_i$.
This procedure gives not only the best determination of $a$ and $b$ 
but also intervals of confidence for these constants.

\begin{figure}
\resizebox{\hsize}{!}{\includegraphics{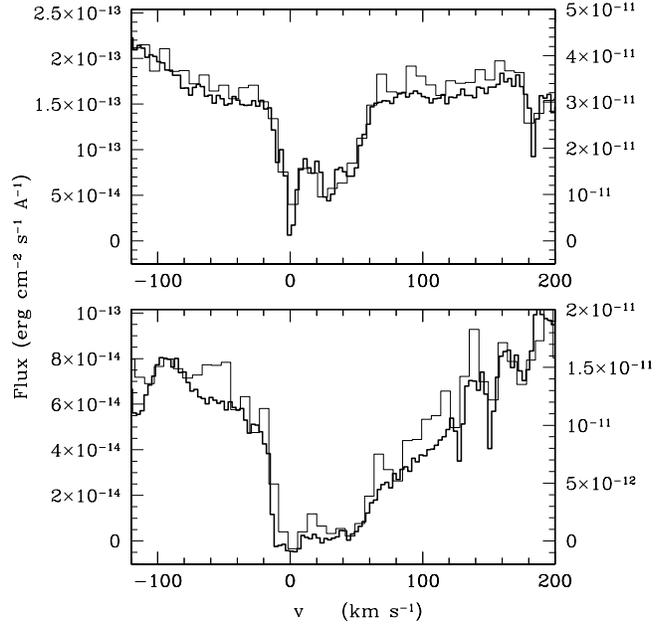}}
\caption[]{
Plot of the disk spectrum (thin line, left axis) and the star spectrum
(thick line, right axis) in the region of the \feii\ lines at 
2374.461~\AA\ (top panel) and 2382.765~\AA\ (bottom panel).
The velocities are relative to \bp\ (21~\kms\ heliocentric).
The data on the disk have been rebinned by 5 pixels, the data
on the star by 2 pixels. The ratio of the right axis to the left axis 
is 197. 
}
\label{plot_feii}
\end{figure}

\subsection{Results in the \feii\ lines}
\label{The feii lines}

\begin{figure}
\resizebox{\hsize}{!}{
\includegraphics{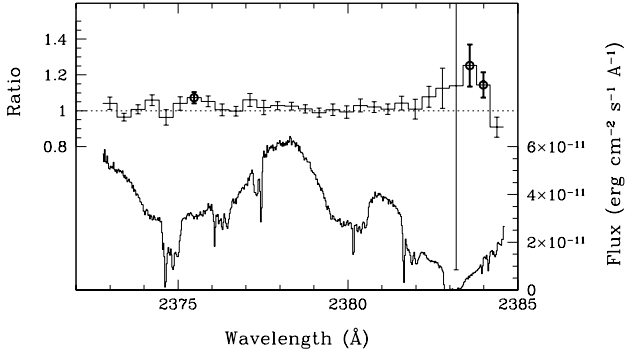}}
\caption[]{Ratio of the disk spectrum to the star spectrum (top histogram, 
left axis) and plot of the star spectrum~\#2 (bottom spectrum, right
axis). The pixels have been rebinned in boxes of 50~\kms width.
To evaluate the ratio, the disk spectrum has been rescaled with $a^{-1}=197$ 
and $b=3\cdot 10^{-15}$~\erg .
The errorbars are at 1$\sigma$ level. The points from which a signal is
detected at 2$\sigma$ are plotted in bold.
The dip in the spectrum between 2377~\AA\ and 2378~\AA\ is due to hot diodes.
An excess emission is clearly detected in the red part of the two
strongest \feii\ lines.
}
\label{ratio_feii}
\end{figure}

\begin{figure}
\resizebox{\hsize}{!}{
\includegraphics{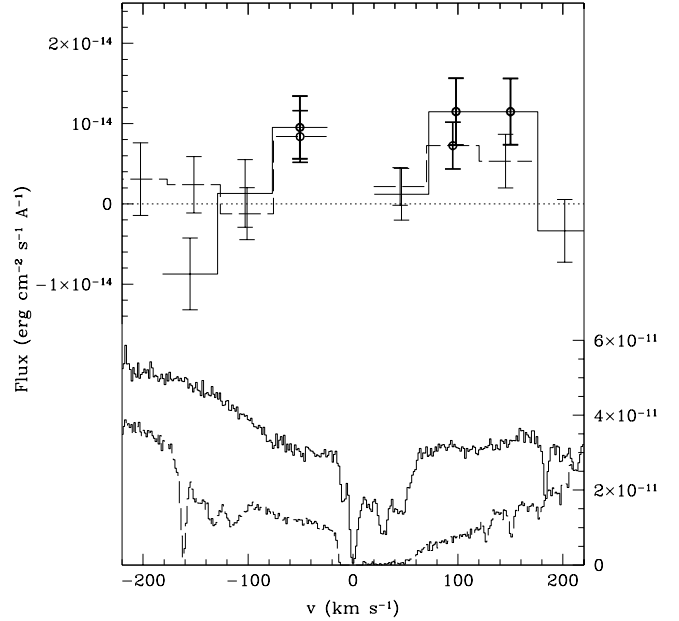}}
\caption[]{
Plot of the \feii\ emission in the disk spectrum at 0.5\arcsec\ obtained 
by the subtraction of the star scattered 
light (top histogram, solid line for the 2374.461~\AA\ \feii\ line and 
dashed line for the 2382.765~\AA\ \feii\ line, left axis).
The data have been rebinned in boxes of 50~\kms\ wide.
Abscissa are radial velocities relative to the radial 
velocity of the star, i.e. heliocentric minus 21~\kms . 
The errorbars are at 1$\sigma$ level.
The points for which a signal is detected at 2$\sigma$ are plotted in bold.
The two bottom spectra are \bp\ spectra (right axis).
Because of the strong \feii\ absorption lines of the stable disk
(at 0~\kms\ relative to \bp , 21~\kms\ heliocentric), pixels with
a velocity between $-$25~\kms\ and $+$20~\kms\ relative to \bp\ are very noisy
and are not taken into account. 
The two \feii\ lines give about the same emission although their shapes
and their levels are very different.
}
\label{vel_feii}
\end{figure}

Spectra in the \feii\ wavelength range were obtained less than
10 minutes apart with the same instrument setting.
The star and the disk spectra can be easily superimposed as
shown in Fig.~\ref{plot_feii}.
An excess of flux clearly appears 
in the blue {\it and} the red part 
of the two strongest \feii\ lines at rest
wavelength 2374.461~\AA\ and 2382.765~\AA .
If $a$ and $b$ are determined as described in Sect~\ref{method}, 
we get $a^{-1}=197\pm 3$ 
and $b=(3 \pm 2)\cdot 10^{-15}$~\erg.
With these parameters, a plot of the ratio of the disk to the star spectra 
reveals an excess emission 
at 2$\sigma$ level (Fig.~\ref{ratio_feii}). 
The flux from the disk 
can be evaluated by the difference between the two 
spectra ($F_{obs}(\lambda)-(a F_*(\lambda) + b)$), 
and is $F_{disk}\sim 10^{-14}$~\erg\ between 
50 and 150 \kms\ in the red part of the lines
and around $-$50~\kms\ in the blue part
(Fig.~\ref{vel_feii}).
This emission is about 1~\AA\ wide and stronger in the red part of the lines.

\subsection{Discussion}

The detection of apparent excess emission in the two \feii\ lines
can be explained in different ways.
We propose that this can be a real detection of the 
emission through the scattering of the starlight by the \feii\ ions
in very high velocity motions.
This will be discussed in detail in the next section.
However other possibilities must be evaluated.

First, this feature obviously cannot be simply due to a possible miscalculation
(underestimate) of $b$, the background correction. 
Indeed, by underestimating $b$, we could find false emission-like features
at wavelength where the flux is low. 
But, one should increase the estimate of $b$ 
above its 4$\sigma$ upper limit ($8\cdot 10^{-15}$~\erg )
to explain the detected emission feature only by this effect.
In addition, the emission is detected in both \feii\ lines, 
in particular in the \feii\ line at 2374.461~\AA\ where the level 
is far above the zero level, and for which an error on the
estimate of $b$ has almost no effect on the result.
In fact, $b$ is well-determined by
the very bottom of the strongest \feii\ lines where the level is clearly 
less than $1\cdot 10^{-14}$~\erg\ (Fig.~\ref{plot_feii}).

However, apparent emissions due to the addition of 
the statistical noise and a bad estimate of the background level
might be possible.
But, although not excluded, it is very unlikely
that 
this coincidence 
can contribute to give apparent emissions in the two sides of
the two strongest \feii\ lines.

The most important alternative to emission by \feii\ is
a time variation of the \bp\ spectrum 
between the observations of the template (the star) and the disk spectra. 
If the absorption component in the \feii\ lines had significantly decreased 
during the acquisition of the data, then the result is an apparent
excess of emission in the second spectrum obtained with
the slit off the star. 
Although the time between both spectra has been minimized,
this possibility cannot be excluded without new observations,
for example on the other side of the disk
where the emission should be stronger in the blue.

\section{The \feii\ emission lines}
\label{The feii emission lines}

\subsection{The dynamics of the \feii\ ions}

If this detection is really due to emission by \feii\ ions, 
the lines width must be explained through
the dynamics of these \feii\ ions in the disk.
The \feii\ ions must be ejected from the \bp\ system by the 
radiation pressure which is stronger than the gravitation 
by a factor $\beta_{FeII}\approx 5$ (Lagrange et al., 1998).
After ejection, they rapidly reach a constant asymptotic velocity $v_{\infty}$.
If they are ejected from a body on a circular orbit,  
$v_{\infty}\sim  \sqrt{(2\beta-1)(GM/a_0)}$,
where $a_0$ is the radius of the orbit of the parent body.
If they are ejected from a comet on a parabolic orbit, the 
final velocity is $v_{\infty}\sim  \sqrt{(2\beta GM/a_0)}$.
In this simple scheme, the observed final 
velocity of about 100~km~s$^{-1}$ (Fig.~\ref{vel_feii})
corresponds to a production at about 1.5~AU from the star,
or similarly to the absence of gas drag beyond that distance.

\begin{figure}
\resizebox{\hsize}{!}{\includegraphics{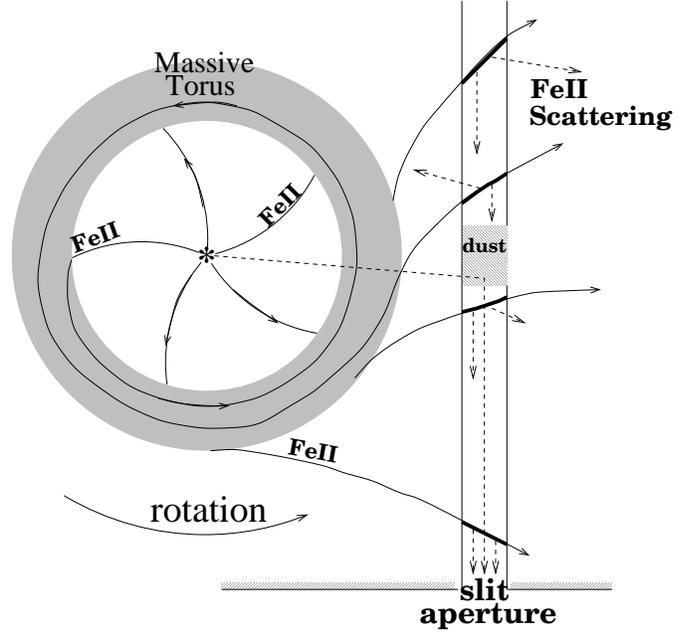}}
\caption[]{Sketch of the possible dynamics of the \bp\ disk.
FeII ions are produced close to the star, spiral into a
massive torus at about 1~AU and 
then cross the line of sight defined by the slit.
We see the resonant scattering through emission 
lines blueshifted and redshifted by the radiation pressure. 
The observed larger amount of redshifted \feii\ is due either to 
the rotation of the massive torus which drags these \feii\ ions,
or to an asymmetrical production of ions as inferred from the 
falling comets 
scenario.
In summary, the large velocities
are due to the radiation pressure, and the asymmetry
between the red and the blue is due to the rotation of the system.
}
\label{sketch}
\end{figure}

The emission lines 
are stronger in the red than in the blue.
This is similar to the asymmetry already observed in the cometary absorption
lines which are mainly redshifted (Beust et al. 1996). 
This last asymmetry is well-explained by
the evaporation of comets with a small range of longitude of 
periastron (Beust et al. 1998). 
An alternative explanation for the observed asymmetry in the emission lines 
could be an extended shape for the dragging torus of gas
needed to support the radiation pressure 
on the \caii\ and \feii\ ions observed at zero radial velocity in the stable
gaseous disk (Lagrange et al. 1998, Beust et al. 1998).
For both explanations, 
it is clear that the observed south-west branch of the disk 
must be the ``red side'' of the disk (Fig.~\ref{sketch}). 
This provides an observational test : 
the north-east branch must present a larger emission in the blue lines.

\subsection{The \feii\ density}

The total brightness of \feii\ emission lines can be evaluated to be:
\begin{equation}
 F_{emission} = \frac{\Omega d^2 s F_\nu^{\beta {\rm Pic}}}{4\pi}  \int n(r)/r^2 dx
\end{equation}
in ${\rm erg}~{\rm cm}^{-2}~{\rm s}^{-1}$, where 
$\Omega$ is the solid angle covered by the spectrograph slit.
The SSA slit (0.22\arcsec x0.22\arcsec ) gives $\Omega = 10^{-12}$.
$d$ is the distance to \bp\ ($d = 19.3$~pc$ \approx 6\cdot  10^{19}$cm).
$s$ is the frequency integral of the cross section.
$F_\nu^{\beta {\rm Pic}}$ is the brightness per unit of frequency 
of \bp\ seen from the Earth at the relevant wavelength.
$n(r)$ is the density of the observed ion at a distance $r$ 
from the central star. $dx$ is the differential length along the line of sight.
We can define a weighted integral equivalent to the column density by 
\begin{equation}
\tilde N_{r_0} \equiv \int\frac{n(r)}{r^2}r^2_0 dx 
= \frac{4\pi r_0^2 F_{emission}}{\Omega d^2 s F_\nu^{\beta \rm{Pic}}},
\end{equation}
where $r_0$ is the impact parameter of the line of sight
($r_0(0.5\arcsec )=10$~AU).
For the \feii\ 2382 \AA\ line, 
$s= 8\cdot  10^{-3}$~cm${^2}$~s$^{-1}$
and the
observed emission is 
$
F_{emission}=F_{\lambda}^{disk} \Delta \lambda 
        \sim 10^{-14} {\rm erg}~{\rm cm}^{-2}~{\rm s}^{-1}.
$
Finally, we have
\begin{equation}
\tilde N_{10{\rm AU}}\cong 5.2\cdot 10^{26}
 \left(\frac{F_{emission}}{\rm erg\ cm^{-2}\ s^{-1}}\right) {\rm cm}^{-2}
\end{equation}
\begin{equation}
\tilde N_{10{\rm AU}}
\sim 5\cdot 10^{12} {\rm cm}^{-2}
\end{equation}
This value is consistent with the \feii\ column density
($N_{\feii} = \int n(r) dr = 3 \cdot 10^{14}$~cm$^{-2}$) and the 
hypothesis that \feii\ should be gathered in the dragging torus around 1~AU
and that they 
have an $r^{-2}$ distribution beyond this torus (Lagrange et al. 1998).
To evaluate the \feii\ volume density, $n_0$, at $r_0$, 
we can define the dimensionless quantity 
$K$ by 
$
K \equiv \int n(r) r_0 /(n_0 r^2) dx.
$
Then, 
\begin{equation}
n_0=\frac{4\pi r_0 F_{emission}}{\Omega d^2 s F_\nu^{\beta {\rm Pic}} K}
=\frac{\tilde N}{r_0 K}.
\end{equation}
If we make the assumption that $n(r)=n_0 (r_0/r)^\alpha$, 
we have $K_{\alpha=0}= \pi$, $K_{\alpha=1}=2$, 
$K_{\alpha=2}=\pi/2$, ..., $K_{\alpha=8} = 35\pi/128$, {\it et cetera}. 
$\alpha=2$ would correspond to gas expelled by radiation pressure.
For reasonable value of $\alpha$, within a factor of 2, $K\sim 2$.
As a final result, we get 
\begin{equation}
n_0 \sim 2\cdot 10^{-2} {\rm cm}^{-3},\ \ \ {\rm at}\ \sim 10\ {\rm AU}.
\end{equation}

The calculation has been done with the hypothesis of an optically thin line.
We also assume that the filling factor in the vertical direction is~1, 
thus the obtained density $n_0$ is in fact a lower limit.  
The other \feii\ line at 2374~\AA\ has an oscillator strength 10~times smaller
but the stellar flux is larger at this wavelength 
($\sim 4\cdot 10^{-11}$~\erg ).
It gives about the same value of density within a factor of~2.

The corresponding \feii\ production rate 
can be roughly estimated by the flux of material 
($ v=100$~km~s$^{-1}$) through the area observed at 10~AU 
with high $H(0.22\arcsec )=4.4$~AU:
$
Q (\feii)
= 2 \pi r_0 H  v n_0
\approx 10^9$~kg~s$^{-1}$.
With solar abundances, this
corresponds to the total disruption of about 10~asteroids per 
year with a radius of 30~km.

\section{Need for confirmation}
\label{confirmation}

To confirm the tentative detection presented in this paper, new
observations are really needed. 
The first obvious method will be to observe the other (north-east) side of 
the disk which should present the same feature except blueshifted
instead of redshifted.
We can also observe other lines which should present the
same characteristics.
The \feii\ lines at 2600~\AA\ and \mgii\ lines at 2800~\AA\ are 
well-suited candidates.
The \crii\ line at 2050~\AA\ with an intermediate
$\beta_{\crii}\approx 3$ is also an interesting target
(Lagrange et al. 1998).
STIS, with long-slit capability is technically better suited to 
this observation.
 
\section{Conclusion}
\label{Conclusion}

HST observations planned to detect the emission lines
of ions in the \bp\ disk with the spectrograph slit placed off the star
gave a marginal detection of 
possible
emission lines from \feii\ ions
at 0.5\arcsec\ from \bp .
This would suggest an \feii\ density of $\sim 2\cdot 10^{-2}$cm$^{-3}$ 
at 10~AU. 

If real, this indicates
a large production rate of gas and dust 
equivalent to the total disruption of 
10 bodies of 30 kilometers in radius per year.
This corresponds to a production rate 
of about $2\cdot 10^{-7}$M$_{\rm Earth}$ per year.
New observations are obviously needed to confirm 
this detection.

%
%

\end{document}